\documentclass[11pt,dvips]{article}
\usepackage{epsfig,times} 
\usepackage{picinpar}
\usepackage{wrapfig}
\usepackage{floatflt}
\makeatletter
\renewcommand{\section}{\@startsection{section}{1}{0in}
	{0.4\baselineskip}{0.1\baselineskip}{\Large\bf}}
\renewcommand{\subsection}{\@startsection{subsection}{2}{0in}
	{0.25\baselineskip}{-\baselineskip}{\large\bf}}
\renewcommand{\subsubsection}{\@startsection{subsubsection}{3}{0in}
	{0.1\baselineskip}{-\baselineskip}{\normalsize\bf}}

\makeatother
\begin{document}

%
\thispagestyle{myheadings}
%
\markright{OG 2.1.37}
\begin{center}
%
{\LARGE \bf Whipple Observations of BL Lac Objects}
\end{center}

\begin{center}
%
%
{\bf M. D'Vali$^{1}$, I.H. Bond$^{1}$, S.M. Bradbury$^{1}$, J.H. Buckley$^{2}$
A.M. Burdett$^{1,3}$, D.A. Carter-Lewis$^{4}$, M. Catanese$^{4}$, M.F. Cawley$^{5}$, D.J. Fegan$^{6}$, S.J. Fegan$^{3}$, J.P. Finley$^{7}$, 
J.A. Gaidos$^{7}$, T.A. Hall$^{7}$, A.M. Hillas$^{1}$, J. Knapp$^{1}$, 
F. Krennrich$^{4}$, S. Le Bohec$^{4}$, R.W. Lessard$^{7}$, C. Masterson$^{6}$, 
S.D. Myles$^{1}$, J. Quinn$^{6}$, H.J. Rose$^{1}$, F.W. Samuelson$^{4}$, G.H. Sembroski$^{7}$, 
V.V. Vassiliev$^{3}$, T.C. Weekes$^{3}$}\\
{\it $^{1}$Department of Physics \& Astronomy, University of Leeds, Leeds, LS2 9JT, UK\\
$^{2}$Department of Physics, Washington University, St. Louis, MO 63130, USA\\ $^{3}$Fred Lawrence Whipple Observatory, Harvard-Smithsonian CfA, P.O. Box 97, Amado, AZ 85645-0097, USA\\ $^{4}$Department of Physics and Astronomy, Iowa State University, Ames, IA 50011-3160, USA\\ $^{5}$Physics Department, St. Patrick's College, Maynooth, County Kildare, Ireland\\ $^{6}$Experimental Physics Department, University College, Belfield, Dublin 4, Ireland\\ $^{7}$Department of Physics, Purdue University, West Lafayette, IN 47907, USA}
\end{center}

\begin{center}
{\large \bf Abstract\\}
\end{center}
\vspace{-0.5ex}
%
%
Only BL Lac objects have been detected as extragalactic sources of
very high energy (E $>$ 300 GeV) gamma rays.  Using the Whipple
Observatory Gamma-ray Telescope, we have attempted to detect more BL
Lacs using three approaches.  First, we have conducted surveys of
nearby BL Lacs, which led to the detections of Mrk 501 and 1ES
2344+514.  Second, we have observed X-ray bright BL Lacs when the RXTE
All-Sky Monitor identifies high state X-ray emission in an object, in
order to efficiently detect extended high emission states.  Third, we have conducted rapid
observations of several BL Lacs and QSOs located close together in the
sky to search for very high flux, short time-scale flare states such
as have been seen from Mrk 421.  We will present the results of a survey 
using the third observational technique. 

%

\vspace{1ex}

\section{Introduction}
\label{intro.sec}
Three BL Lacertae objects (BL Lacs) have been detected at very high energies (VHE, E $>$ 300 GeV) by the Whipple Observatory Gamma-ray Telescope.  The discovery of the first of these objects: Mrk 421 (Punch et al. 1992), prompted a comprehensive search of other BL Lacs, resulting in the detection of Mrk 501 (Quinn et al. 1996) and 1ES 2344+514 (Catanese et al. 1998).  The detection of other BL Lacs could lead to further constraints on gamma-ray emission models and also to an estimate of the density of extragalactic background infra-red radiation through its attenuation effects on VHE gamma-ray spectra (Biller et al. 1998).

BL Lacs belong to a class of active galactic nuclei (AGN) called blazars.  The radio to optical/X-ray emission produced by these objects is dominated by non-thermal synchrotron radiation believed to be from a relativistic jet moving almost directly towards the observer, so that its intensity is greatly enhanced by Doppler boosting.  Gamma-ray flares almost as short as the basic timescale associated with accretion on to a black hole of mass 10$^{8}$ - 10$^{9}$ M$_{\odot}$ have been seen, indicating that we may be observing processes right at the beginning of the jet.  Models of the high energy emission include synchrotron self-Compton (Bloom \& Marscher 1996), inverse Compton scattering of low energy photons from outside the jet (Sikora, Begelman \& Rees 1994), and cascades produced by high energy protons e.g., (Mannheim 1993).

Observations of both Mrk 421 and Mrk 501 have shown that their emission can vary by nearly two orders of magnitude, with extreme variability on timescales from 15 minutes (Gaidos et al. 1996) to years (Quinn et al. 1999).  The fact that Mrk 501 became so bright in 1997 (Protheroe et al. 1997) having been considerably fainter than Mrk 421 in previous years suggests that other BL Lacs not yet detected could do the same, and become detectable.  By monitoring several BL Lacs with nightly 10 minute observations throughout the observing season, we maximise our chances of seeing an episode of high emission.  Table 1 lists two groups of objects observed in this manner during February 1999, when there was a suitable two hour slot in our observing schedule. \\
 
\section{Observations and Analysis}
\label{format.sec}
The VHE observations reported in this paper were made with the 10 m imaging atmospheric \v{C}erenkov telescope (Cawley et al. 1990) located at the Whipple Observatory on Mt. Hopkins in southern Arizona (elevation 2,300 m a.s.l.).  A wide field-of-view camera, consisting of 331 photomultiplier tubes, is mounted in the focal plane of the reflector and has a field of view of $\sim$ 4.8 degrees.  Images of atmospheric \v{C}erenkov radiation from air showers produced by gamma rays and cosmic rays are cleaned and parameterised to allow candidate gamma ray selection as described by Quinn et al. (1999).

The results reported in this paper use a tracking analysis wherein events whose orientations do not point towards the source direction are used to estimate the background event rate.  Approximately 100 non-source data files, consisting of off source observations and non-detected objects were combined to estimate the factor which converts the off-source events to a background estimate.  The count rates are converted to integral fluxes by expressing them as a multiple of the Crab Nebula count rate and then multiplying that fraction by the Crab Nebula flux, I($>$500 GeV) = 5.9 x 10$^{-11}$ photons cm$^{-2}$ s$^{-1}$ (Hillas et al. 1998).  This procedure assumes that the Crab Nebula VHE gamma ray flux is constant, and that the object's photon spectrum is identical to that of the Crab Nebula between 0.5 and 10 TeV, which may not be the case.  If no significant emission is seen from a candidate source, a 99.9\% confidence level upper limit is calculated using the method of Helene (1983).
\begin{figure}[h]
\centerline{\epsfig{file=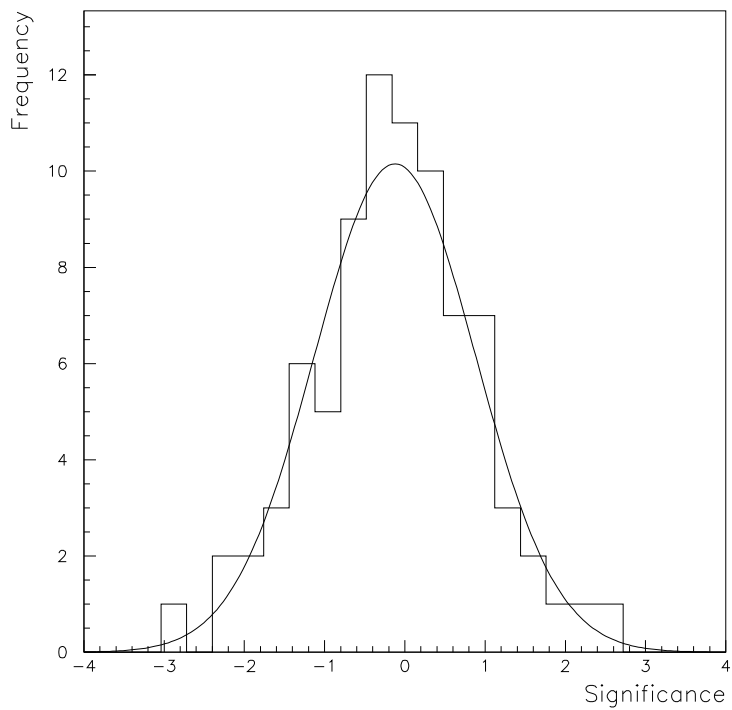,width=4.2in}}
\vspace{-0.5cm}
\caption{Gaussian fit to the significance distribution of the BL Lac object observations.}
\end{figure}

\section{Results}
\label{results.sec}
In Table 1 we present the results of observations from February 1999.  The sources were chosen primarily for their close, high elevation positions in the sky during that period, to minimise time lost slewing between them (which accounted for less than 10\% of the total observing time).  There is no evidence of emission from any of the objects in this survey, so 99.9\% confidence level upper limits have been determined for each source.  The distribution of the significances of the observed excesses is shown in Figure~1.

\begin{table}[h]
\caption[Summary of BL Lac Observations.] 
		{\em Summary of BL Lac Observations.}
\begin{center}
\begin{tabular}{lcccccccc}
\hline\hline
&&&\textbf{Observ.}&\textbf{Exp.}  &&\textbf{Max.}  & \textbf{Flux}\\
\textbf{Object}  &\textbf{z}  & \textbf{Type$^a$}  &\textbf{Epoch} &\textbf{(mins)}  &\textbf{Excess}  &\textbf{Excess$^b$}  &\textbf{(Crab)$^c$}  &\textbf{Flux$^d$}\\
\hline
Mrk 180  & 0.046  & X  & 10-20 Feb. 1999  & 77  & 0.04  & 1.05  & $<$0.663  & $<$3.91\\
W Comae  & 0.102  & R  & 10-23 Feb. 1999  & 296  & -0.37  & 2.03  & $<$0.198  & $<$1.17\\
1ES 1118  & 0.124  & X  & 10-21 Feb. 1999  & 118  & -0.57  & 2.22  & $<$0.257  & $<$1.52\\
1H 1219  & 0.130  & X  & 11-24 Feb. 1999  & 60  & -0.06  & 1.37  & $<$0.426  & $<$2.51\\
RXJ 11365  & 0.135  & X  & 10-20 Feb. 1999  & 80  & -1.23  & 1.62  & $<$0.451  & $<$2.66\\
1ES 1212  & 0.136  & X  & 12-24 Feb. 1999  & 69  & 0.32  & 1.41  & $<$0.591  & $<$3.49\\
1ES 1255  & 0.141  & X  & 12-24 Feb. 1999  & 96  & 0.11  & 2.48  & $<$0.361  & $<$2.13\\
1ES 1239  & 0.150  & X  & 12-24 Feb. 1999  & 69  & -0.24  & 1.10  & $<$0.535  & $<$3.16\\
1229+645  & 0.164  & R  & 10-24 Feb. 1999  & 154  & 0.51  & 1.10  & $<$0.456  & $<$2.69\\
ON 325  & 0.237  & X  & 11-23 Feb. 1999  & 59  & 1.61  & 1.65  & $<$0.648  & $<$3.82\\
\hline\hline
\multicolumn{9}{l}{$^{a}$Indicates whether the object is a radio selected (R) or X-ray selected (X) BL Lac object.}\\
\multicolumn{9}{l}{$^{b}$Maximum excess per 10 minute observing run.}\\ 
\multicolumn{9}{l}{$^{c}$Upper limit expressed in units of the Crab Nebula flux.}\\
\multicolumn{9}{l}{$^{d}$Integral flux upper limit quoted above 500 GeV in units of 10$^{-11}$cm$^{-2}$s$^{-1}$, at the 99.9\% confidence level.}
\end{tabular}
\end{center}

\label{tab:bllacs}
\end{table}

To demonstrate the validity of this search method, we have taken four 10 minute data samples of Markarian 501 observations from 15 February 1999 when it was in a high state of emission.  Figure 2 shows the gamma ray rates and the corresponding significance of each sample. The fluxes are expressed in units of the Crab rate above $>$500 GeV.  Note that at a rate of 2 Crab an excess of $\sim$ 3$\sigma$ can be seen in 10 minutes.  A detailed analysis of the flux variability of Markarian 501 can be found in Quinn et al. (1999), and a report on 1999 Mrk 501 observations will appear in another paper to be presented at this conference (Breslin et al. 1999).
\begin{figure}[h]
\vspace{-1cm}
\centerline{\epsfig{file=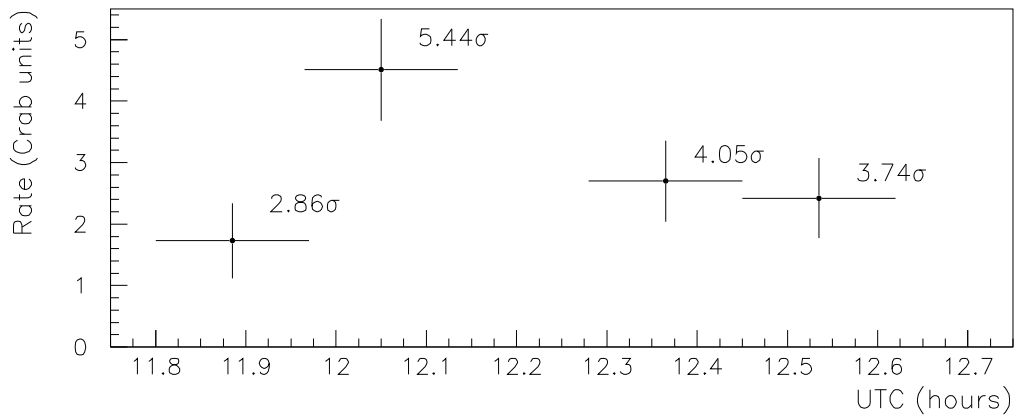,width=15cm}}
\vspace{-0.5cm}
\caption{Gamma ray rate expressed in units of the Crab rate for Markarian 501 observations taken on 15 February 1999, illustrating that a BL Lac in a high flux state can be detected in under 10 minutes.}
\end{figure}

\pagebreak

\section{Conclusions}
\label{conclusions.sec}
The results of the Markarian 501 observations show that a signal can be detected from a BL Lac object during a high flux state in only 10 minutes.  By scanning these closely grouped sources for 10 minutes each we are able to observe a high proportion of very high energy BL Lac candidates on a regular basis within a given amount of observing time.  Further observations need to be made using this technique; in view of the fact that the frequency of very bright emission episodes from BL Lacs is unknown, our initial non-detection is in no way discouraging.\\    

\section{Acknowledgements}
\label{acknowledgements.sec}

We acknowledge the technical assistance of K.Harris and E.Roache. This
research is supported by the U.S. Department of Energy, by PPARC (U.K.) and
Enterprise-Ireland. \\

\vspace{1ex}
\begin{center}
{\Large\bf References}
\end{center}
Biller, S.D., et al. 1998, Phys. Rev. Lett., 80, 2992 \\
Bloom, S.D., \& Marscher, A.P. 1996, ApJ, 461, 657 \\
Breslin, A., et al., 1999, Proc. 26th ICRC (Salt Lake City, 1999) OG.2.1.23 \\
Catanese, M., et al. 1998, ApJ, 501, 616 \\
Cawley, M. F., et al., 1990, Exp. Astron., 1, 173 \\
Gaidos, J., et al., 1996, Nature, 383, 319 \\
Helene, O., 1983, Nucl. Instr. Meth., 212, 319 \\
Hillas, A. M., et al., 1998, ApJ, 503, 744 \\
Mannheim, K., 1993, A\&A, 269, 67 \\
Protheroe, R., et al., 1997, Proc. 25th ICRC (Durban), 8, 317 \\
Punch, M., et al., 1992, Nature, 358, 477 \\
Quinn, J., et al., 1996, ApJ, 456, L83 \\
Quinn, J., et al., 1999, ApJ, 518 in press \\ 
Sikora, M., Begelman, M. C., \& Rees, M. J., 1994, ApJ, 421, 153 \\

\end{document}